\documentclass[a4paper,12pt,tightenlines,notitlepage,nofootinbib, superscriptaddress]{revtex4-2}
\usepackage{amsmath,amssymb}
\usepackage{graphicx}
\usepackage{newtxtext}
\usepackage{newtxmath}
\usepackage[x11names]{xcolor}
\usepackage[unicode,colorlinks,citecolor=Blue3,linkcolor=Blue3,bookmarks=true]{hyperref}

\newcommand{\ie}{{\it i.e.\,}}
\newcommand{\eg}{{\it e.g.\,}}

\newcommand{\hence}{\ \Rightarrow\ }
\newcommand\bra[2][]{#1\langle {#2} #1|}
\newcommand\ket[2][]{#1|{#2} #1\rangle}
\newcommand{\braket}[2]{ \langle #1 | #2 \rangle}
\renewcommand{\Re}{\mathop{\rm Re}\nolimits}
\DeclareMathOperator{\Tr}{Tr}

\begin{document}

	\title{Using non-Gaussian quantum states for detection of a given phase shift}

	\author{V.\,L.\,Gorshenin}
	\email{valentine.gorshenin@yandex.ru}
	\affiliation{Russian Quantum Center, Skolkovo 121205, Russia}
	\affiliation{Moscow Institute of Physics and Technology, 141700 Dolgoprudny, Russia}
	\author{F.\,Ya.\,Khalili}
	\affiliation{Russian Quantum Center, Skolkovo 121205, Russia}

\begin{abstract}

Injecting a non-Gaussian (Fock or Shr\"odinger cat) quantum state into the dark port of a two-arm interferometer and a strong classical light into the bright one, it is possible, in principle, to detect a given phase shift unambiguously using the orthogonality between the original and displaced in the interferometer non-Gaussian states.

The optical losses degrade the sensitivity, introducing the finite "false positive" and "false negative" detection errors. However, using the state-of-art photodetectors, it is still possible to obtain better detection fidelity than in the case of Gaussian quantum states.

\end{abstract}

	%
	%
	%
	%
	%

	\maketitle

\section{Introduction}\label{sec:Intro}

A standard task of optical interferometry is a measurement of an unknown phase shift. The corresponding quantum sensitivity limitations were explored in detail in numerous works, see \eg the reviews \cite{Demkowicz_PIO_60_345_2015, Andersen_ch35_2019, 22a1SaKh} and the references therein. In particular, it is known that in the baseline case of the coherent quantum state of the probing light, the mean squared error of the phase measurement corresponds to the shot noise limit:
\begin{equation}\label{SNL}
  \Delta\phi_{\rm SNL} = \frac{1}{2\sqrt{N}} \,,
\end{equation}
where $N$ is the mean number of photons used for the measurement. This sensitivity could be improved without increase of $N$ by using squeezed quantum states, as it was proposed by C.\,Caves in his pioneering work \cite{Caves1981}. He showed that in the moderate squeezing case, $e^{2r}\ll N$, where $r$ is the logarithmic squeeze factor, the phase measurement error can be reduced by the factor $e^r$:
\begin{equation}\label{dphi_sqz}
  \Delta\phi_{\rm SQZ} = \frac{e^{-r}}{2\sqrt{N}} \,.
\end{equation}
This method is now used in the laser gravitational-wave detectors \cite{Nature_2011, Nature_2013, Grote_PRL_110_181101_2013, Dwyer_Galaxies10_020046_2022}. In case of very strong squeezing, $e^{2r}\sim N$, the sensitivity approaches the so-called Heisenberg limit $\sim1/N$ (see \eg Ref.\,\cite{17a1MaKhCh}).

The possible use of more exotic ``truly quantum'' non-Gaussian states, the ones that have been described by the Wigner quasi-probability functions \cite{Schleich2001, Schleich2001} having a non-Gaussian shape, was also explored in literature, see \eg Refs.\,\cite{Holland_PRL_71_1355_1993, Lee_JMO_49_2325_2002, Campos_PRA_68_023810_2003, Berry_PRA_80_052114_2009, Pezze_PRL_110_163604_2013, Perarnau-Llobet_QST_5_025003_2020, Shukla_OptQEl_55-460_2023, Shukla_PhOpen_18_100200_2024} and the review \cite{Demkowicz_PIO_60_345_2015}. However, no any decisive advantages of non-Gaussian states were demonstrated in these works. In addition, it was shown in Ref.\,\cite{Lang_PRL_111_173601_2013} that in the most important from the practical point of view case of the two-arm interferometer with the strong classical carrier (\ie the coherent state) fed into the one (bright) port and some quantum state $\ket{\psi}$ --- into another (dark) one, the  optimal choice for $\ket{\psi}$ is a Gaussian state.

At the same time, the non-Gaussian states look very promising for another class of interferometric tasks, namely, the binary discrimination between two possible and known in advance values of the phase shift. A simple example is distinguishing between two dielectrics with the known and slightly different refractive indices. Without limiting the generality, one of these values can be set equal to zero, reducing the task to the detection of a given phase shift.

It is known that the Wigner functions of non-Gaussian (and only non-Gaussian) pure states take negative values \cite{Hudson_RMP_6_249_1974}. Taking into account that the scalar product of any two states can be expressed through the convolution of their Wigner functions \cite{Schleich2001}, this means that two quantum states can be orthogonal to each other if at least one of them is a non-Gaussian one. It is also known that quantum states that are orthogonal to each other (and only such states) can be discriminated unambiguously \cite{HelstromBook}. Therefore, non-Gaussian quantum states should allow to detect a given phase shift with very high fidelity, limited, in principle, only by optical losses.

Of all kinds of non-Gaussian states, two are the most interesting from a practical point of view, because the technologies of their preparation are more or less developed, namely the Fock states and the Schr\"odinger cat (SC) states.

In Ref.\,\cite{Gorshenin_LPL_21_065201_2024}, the scheme of unambiguous detection of a given phase shift using the SC state was explored. In the current paper, we extend the analysis of \cite{Gorshenin_LPL_21_065201_2024} in the following directions: (i) in addition to the SC states, we consider also the Fock states;  (ii) following Ref.\,\cite{Caves1981}, we consider the squeezing of the non-Gaussian state at the input of the interferometer and the anti-squeezing at the output; (iii) we take into account the non-ideal quantum efficiency of the photodetectors.

The paper is organized as follows. In Sec.\,\ref{sec:ifos} we introduce the interferometric scheme considered in this work. In Sec.\,\ref{sec:fock} calculate its sensitivity, assuming that the Fock state is injected into the interferometer dark port. In Sec.\,\,\ref{sec:sc}, we consider the case of the Schr\"odinger cat state. In Sec.\,\ref{sec:conclusion} we discuss the prospects of the experimental implementation of the concept we consider here.

\section{Evolution of the light in the interferometer}\label{sec:ifos}

\begin{figure}
  \includegraphics[scale=1.2]{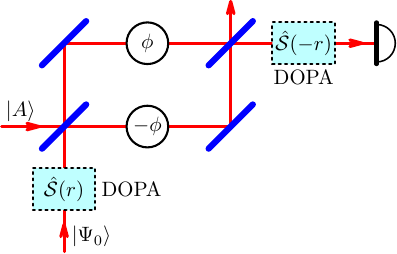}\quad
  \includegraphics[scale=1.2]{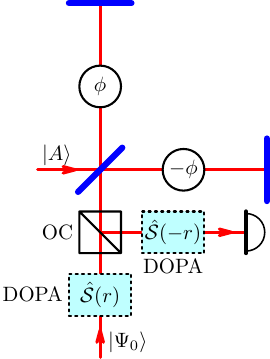}
  \caption{Two equivalent implementations of the non-Gaussian interferometer: the Mach-Zehnder (left) and the Michelson (right). OC --- optical circulator; DOPA --- optional degenerate optical parametric amplifiers; $\ket{A}$ --- the bright coherent state; $\ket{\Psi_0}$ --- a non-Gaussian state.}\label{fig:scheme}
\end{figure}

Consider a standard two-arm (Mach-Zehnder or Michelson) interferometer, see Fig.\,\ref{fig:scheme}. We assume that symmetric beamsplitters are used in this scheme, and the signal phase shifts $\pm\phi$ are introduced antisymmetrically into both arms. This option is immune to the common phase shift and therefore more tolerant to technical noises; due to this, it is used in the high-precision measurements, \eg in the laser GW detectors.

Suppose that it is pumped through one of its input ports (the bright one) by classical (coherent) light, and some quantum state $\ket{\Psi_{\rm in}}$ is injected into the second input port (the dark one). Suppose also that the interferometer is tuned in such a way that if $\phi=0$, then these quantum states are reproduced at the respective bright and dark output ports (the dark fringe regime). Without limiting the generality, we assume that the classical field amplitude $A$ is a real quantity.

A corresponding input/output relations were calculated in  Ref.\,\cite{Gorshenin_LPL_21_065201_2024}. It was shown that in the linear
in small phase shift $\phi$ and quantum fluctuations approximation, the optical field at the bright output port does not depend on $\phi$, whereas the quantum state at the dark output port can be presented as follows:
\begin{equation}\label{ifo_io}
	\ket{\Psi_{\rm out}} = \hat{\mathcal{D}}_{\rm IFO}\ket{\Psi_{\rm in}} \,,
\end{equation}
where
\begin{equation}\label{D_delta_IFO}
	\hat{\mathcal{D}}_{\rm IFO} = e^{iA(\hat{a} + \hat{a}^\dag)\phi} \,,
\end{equation}
is a displacement operator and $\hat{a}$ is a annihilation operator, corresponding to the optical field at the dark port.

Besides the antisymmetric variant of Fig.\,\ref{fig:scheme}, the asymmetric one with strongly unbalanced beamsplitters and phase shift introduced only into one arm was also discussed in Ref.\,\cite{Gorshenin_LPL_21_065201_2024}. It was shown that it is described by the same Eqs.\,\eqref{ifo_io},\,\eqref{D_delta_IFO}, but with $A$ being the amplitude of the classical field at the phase shifting object. Note that in both cases,

\begin{equation}
  A = \sqrt{N} \,,
\end{equation}
where $N$ is the number of photons at the phase shifting object(s). Therefore, our analysis will be valid for both configurations.

Suppose that the quantum state $\ket{\Psi_{\rm in}}$ is prepared by means of squeezing the initial quantum state $\ket{\Psi_0}$:
\begin{equation}\label{Psi_in}
  \ket{\Psi_{\rm in}} = \hat{\mathcal{S}}(r)\ket{\Psi_0} \,,
\end{equation}
where $\hat{\mathcal{S}}(r)$ is the squeeze operator defined by the equation
\begin{equation}\label{sqz}
  \hat{\mathcal{S}}^\dag(r)\hat{a}\hat{\mathcal{S}}(r)
  = \hat{a}\cosh r + \hat{a}^\dag\sinh r \,,
\end{equation}
where $r$ is the logarithmic squeeze factor. Suppose also that the quantum state at the interferometer output is proportionally anti-squeezed, giving the final output state equal to
\begin{equation}
  \ket{\Psi_\delta} = \hat{\mathcal{S}}(-r)\ket{\Psi_{\rm out}}
     = \hat{\mathcal{S}}^\dag(r)\ket{\Psi_{\rm out}} \,.
\end{equation}
Using Eqs\,\eqref{Psi_in}, \eqref{sqz}, it is easy to show that
\begin{equation}\label{Psi_delta}
	\ket{\Psi_\delta} = \hat{\mathcal{D}}(\delta)\ket{\Psi_0} \,,
\end{equation}
where
\begin{equation}\label{D_delta}
  \hat{\mathcal{D}}(\delta)
  = \hat{\mathcal{S}}^\dag(r)\hat{\mathcal{D}}_{\rm IFO}\hat{\mathcal{S}}(r)
	= e^{i\delta(\hat{a} + \hat{a}^\dag)} \,,
\end{equation}
and
\begin{equation}\label{delta}
	\delta = A\phi e^r \,.
\end{equation}
Therefore, the squeezing/antisqueezing sequence increases the phase signal by the factor $e^r$ without affecting the shape of the output quantum state.

It follows from Eq.\,\eqref{Psi_delta} that the problem of detection of a given phase shift $\phi$ reduces to the problem of distinguishing between the quantum states $\ket{\Psi_0}$ and $\ket{\Psi_\delta}$. It was shown in the monograph \cite{HelstromBook}, that the corresponding error probability is equal to
\begin{equation}
  P_{\rm err} = \frac{1}{2}
    \biggl(1 - \sqrt{1 - 4p_0p_\delta|\braket{\Psi_0}{\Psi_\delta}|^2}\biggr) ,
\end{equation}
where $p_0$, $p_\delta$ are {\it a priory} probabilities of, respectively, the absence and presence of the phase shift. It follows from this equation, that orthogonal (and only orthogonal) quantum states can be discriminated unambiguously:
\begin{equation}\label{gen_orthog}
  \braket{\Psi_0}{\Psi_\delta} = 0 \hence P_{\rm err} = 0 \,.
\end{equation}
It was shown also in \cite{HelstromBook}, that in the case of \eqref{gen_orthog}, the optimal measurement procedure for the output optical field is described by the following positive operator-valued measure (POVM):
\begin{equation}\label{POVM}
	\bigl\{\ket{\Psi_0}\bra{\Psi_0}, \ket{\Psi_\delta}\bra{\Psi_\delta}\bigr\} \,.
\end{equation}

\section{Fock states}\label{sec:fock}

\subsection{No optical losses}

Let
\begin{equation}\label{Psi_0_n}
  \ket{\Psi_0} = \ket{n} \,,
\end{equation}
where $\ket{n}$ is the $n$-photon Fock state. In this case,
\begin{equation}\label{Psi_delta_n}
  \ket{\Psi_\delta} = \hat{\mathcal{D}}(\delta)\ket{n} \,,
\end{equation}
see Eq.\,\eqref{Psi_delta}. It is shown in Appendix \ref{app:Fock_orthog}, the scalar product of these states is equal to
\begin{equation}\label{Fock_orthog}
   \braket{\Psi_0}{\Psi_\delta} = \bra{n}\hat{\mathcal{D}}(\delta)\ket{n}
   = L_n(\delta^2)e^{-\delta^2/2} \,,
\end{equation}
where $L_n(\cdot)$ is the $n$-th Laguerre polynomial. Therefore, if $\delta^2$ is equal to one of the root of $L_n$, then the states $\ket{\Psi_0}$ and $\ket{\Psi_\delta}$ can be distinguished with the vanishing error probability $P_{\rm err}=0$. This concept was demonstrated experimentally using the motional Fock states of a trapped ion \cite{Wolf_NComm_10_2929_2019}.

Evidently, the best sensitivity is provided by the smallest (that is the first) root $R_n$. It follows from Eq.\,\eqref{delta}, that in this case, the values $\phi_0$ of the phase shift that can be unambiguously detected are defined by
\begin{equation}\label{phi_Fock}
  |\phi_0| = \sqrt{R_n}\frac{e^{-r}}{\sqrt{N}} \,.
\end{equation}
Note that the first root of the first Laguerre polynomial	is equal to $R_1=1$, and for the higher order polynomials, the first root values gradually decrease with the increase of $n$.

It is easy to see the similarity between Eqs.\,\eqref{phi_Fock} and \eqref{dphi_sqz}. In
both cases, the right-hand side is inversely proportional to the square root of the photons number $N$ and, in addition, could be reduced using the squeezing. However, these quantities
have different meanings: the latter one defines the mean squared value of the measurement
error, while former one is equal to the phase shift that can, in principle, be detected without any error.

\begin{figure*}
	\centering
	\includegraphics[width=1\linewidth]{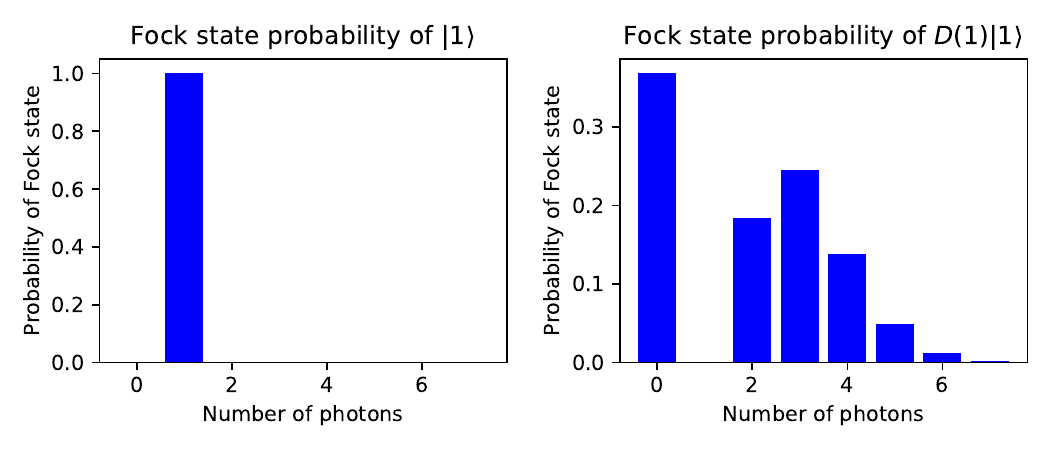}
	\caption{Fock state probability distributions for the states \eqref{Psi_0_n} (left) and \eqref{Psi_delta_n} (right).}\label{fig:fock-state-distribution}
\end{figure*}

In the case of \eqref{Psi_0_n}, it follows also from the orthogonality of $\ket{\Psi_0}$ and $\ket{\Psi_\delta}$, that the probability to detect $n$ photons in the state $\ket{\Psi_\delta}$ is equal to zero. Therefore, instead of the sophisticated generic measurement procedure \eqref{POVM}, the photons number measurement can be used, providing the same sensitivity. In this case, detection of $n$ photons will correspond to the absence of the signal, and of any other photon number --- to the presence. To illustrate this reasoning, we plotted the photon number probability distributions for the states $\ket{\Psi_0}$ and $\ket{\Psi_\delta}$ for the particular case of $n=1$, assuming that the condition \eqref{phi_Fock} is satisfied.

\subsection{With optical losses}\label{sec:nonideal}

It is well-known that the sensitivity gain provided by non-classical states of light is vulnerable to the optical losses. In the modern interferometers, the most serious limiting factor is the photodetection inefficiency, see \eg the loss budget analysis in \cite{Nature_2011, Nature_2013}). Here we calculate how this inefficiency affects the non-Gaussian interferometry.

We model the finite quantum efficiency $\eta<1$ of the detector by means of an imaginary beamsplitter with the power transmissivity $\eta$, which mixes the measured field with an auxiliary vacuum field, as it was proposed in Ref.\,\cite{Leonhardt_PRA_48_4598_1993}.

In order to avoid over-cluttering of the equations, we limit ourselves by the simplest but the most important case of the single-quantum state, $\ket{\Psi_0}=\ket{1}$. It can be shown that in this case, in the absence of the phase shift, the final quantum state of the optical field is described by the density matrix
\begin{equation}\label{rho_delta_0}
  \hat{\rho}_0 = \eta\ket{1}\bra{1} + (1-\eta)\ket{0}\bra{0} \,,
\end{equation}
and in the presence of the phase shift --- by the density matrix
\begin{equation}\label{rho_delta}
  \hat{\rho}_\delta
  = \eta\mathcal{D}(\delta')\ket{1}\bra{1}\mathcal{D}^\dag(\delta')
    + (1-\eta)\mathcal{D}(\delta')\ket{0}\bra{0}\mathcal{D}^\dag(\delta') ,
\end{equation}
where
\begin{equation}
  \delta' = \delta\sqrt{\eta} \,.
\end{equation}

It follows from these equations that the optical losses affect the sensitivity in several ways. Firstly, they reduce the effective signal by the factor $\sqrt{\eta}$. However, for the state-of-art detectors with $1-\eta\ll1$, the corresponding degradation of the signal is small and can be compensated by the proportional increase of the pump power or the squeeze factor. Secondly, the non-zero probability of detection of the photon number $n\ne1$ in the state \eqref{rho_delta_0} creates the `false positive'' (or ``false detection'') error probability:
\begin{equation}\label{P_fp_eta}
  P_{\rm f.p.} = 1 - \bra{1}\hat{\rho}_0\ket{1} = 1-\eta \,.
\end{equation}

Finally, the detection inefficiency dilutes the non-Gaussian first term in Eq.\,\eqref{rho_delta} by the coherent (Gaussian) state second one, creating the non-zero ``false negative'' (or ``signal miss'') error probability that corresponds to detection of a single photon in the presence of the phase shift:
\begin{equation}\label{P_fn_eta}
  P_{\rm f.n.} = \bra{1}\hat{\rho}_\delta\ket{1}
  = \bigl[\eta(1 - \delta'{}^2)^2 + (1-\eta)\delta'{}^2\bigr]e^{-\delta'{}^2} \,.
\end{equation}
The minimum of this probability is provided by
\begin{equation}\label{opt_beta_eta}
  \delta'{}^2 = 1 \,,
\end{equation}
which corresponds to
\begin{equation}\label{phi_Fock_eta}
  |\phi_0| = \frac{e^{-r}}{\sqrt{\eta N}} \,.
\end{equation}
Up to the factor $\sqrt{\eta}$, this value coincides with the one for the lossless case \eqref{phi_Fock}. The corresponding value of the ``false negative'' error probability is equal to
\begin{equation}\label{P_fn_eta}
  P_{\rm f.n.} = \frac{1-\eta}{e} \,.
\end{equation}

\section{Schr\"odinger cat states}\label{sec:sc}

\subsection{No optical losses}

In this subsection, we briefly reproduce the results of Ref.\,\cite{Gorshenin_LPL_21_065201_2024}, complementing them by taking into account the input squeezing and output antisqueezing.

Suppose that the input light prepared in the SC state:
\begin{equation}\label{std_cat}
	\ket{\Psi_0} = \frac{1}{\sqrt{K}}(\ket{\alpha} + \ket{-\alpha}) \,,
\end{equation}
where $\ket{\alpha}$ and $\ket{-\alpha}$ are the coherent states,
\begin{equation}
	K = 2(1+e^{-2|\alpha|^2})
\end{equation}
is the normalization factor and we assume that $\alpha$ is a real number. The corresponding output quantum state is equal to (see Eq.\,\eqref{Psi_delta})
\begin{equation}\label{Psi_delta_SC}
  \ket{\Psi_\delta} = \frac{1}{\sqrt{K}}
	 (e^{i\delta\alpha}\ket{\alpha+i\delta} + e^{-i\delta\alpha}\ket{-\alpha+i\delta}) \,,
\end{equation}
It is easy to show, that in this case,
\begin{equation}\label{overlap_SC}
	\braket{\Psi_0}{\Psi_\delta}
	= \frac{2e^{-\delta^2/2}}{K}(\cos2\alpha\delta + e^{-2\alpha^2}) \,.
\end{equation}
Zeros of this function are equal to
\begin{equation}\label{delta_k}
	\delta_k = \frac{\arccos(-e^{-2 \alpha^2}) + 2 \pi k}{2 \alpha} \,,
\end{equation}
where \(k\) is an integer number. Evidently, the best sensitivity is provided by \(k=0\); the phase shift that in this case can be unambiguously detected is equal to (see Eq.\,\eqref{delta})
\begin{equation}\label{phi_optimal}
	\phi_0 = \frac{\arccos(-e^{-2 \alpha^2})}{2\alpha}\times\frac{e^{-r}}{\sqrt{N}} \,,
\end{equation}
compare with Eqs.\,\eqref{dphi_sqz} and \eqref{phi_Fock}.

The numerator of \eqref{phi_optimal} quickly converges to $\pi/2$ with the increase of $\alpha$ (if \(\alpha > 1.5\) then the difference $<1\%$). Therefore, $\phi_0$ can be approximated as follows:
\begin{equation}\label{phi_opt_approx}
	\phi_0 \approx \frac{\pi}{4\alpha\sqrt{N}}e^{-r} \,.
\end{equation}

Unfortunately, it is unclear, how the corresponding optimal measurement procedure, described by the POVM \eqref{POVM}, can be implemented in practice. Therefore, consider a more practical procedure based on the photon number measurement of at the dark output port.

\begin{figure}
	\centering
	\includegraphics[width=0.5\linewidth]{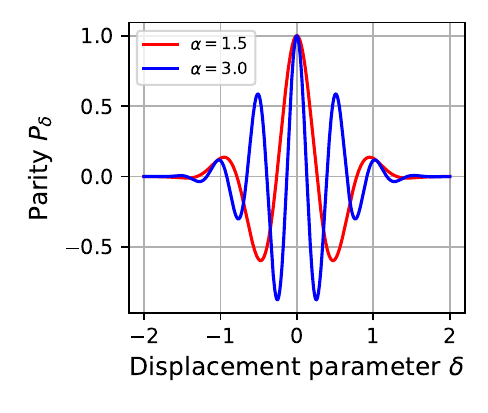}
	\caption{Parity of the displaced SC state, see Eq.\,\ref{parity}) as a function of the displacement parameter \(\delta\) for \(\alpha = 1.5\) (red line) and \(\alpha = 3\) (blue line).}
	\label{fig:par-by-delta-alpha}
\end{figure}

It is easy to show that the initial SC state $\ket{\Psi_0}$ is a superposition of even Fock states only. It is shown also in Ref.\,\cite{Gorshenin_LPL_21_065201_2024} that in the displaced case of $\delta\neq 0$, the odd Fock states appear and, with the increase of $\delta$, become dominant. The corresponding probabilities of obtaining the even and the odd photon numbers are equal to
\begin{equation}\label{p_even_odd}
  p_{\rm even} = \frac{1 + P_\delta}{2}\,,\quad p_{\rm odd} = \frac{1 - P_\delta}{2} \,,
\end{equation}
where
\begin{equation}\label{parity}
	P_\delta = \bra{\Psi_\delta}(-1)^{\hat{n}}\ket{\Psi_\delta}
	= e^{-2\delta^2}\frac{\cos4\alpha\delta + e^{-2\alpha^2}}{1 + e^{-2\alpha^2}}
\end{equation}
is the parity of the state $\ket{\Psi_\delta}$ and $\hat{n}$ is the photon number operator. In Fig.\,\ref{fig:par-by-delta-alpha}, the parity $P_\delta$ is plotted as a function of \(\delta\) for some characteristic values of \(\alpha\).

The following strategy can be used in this case: detection of an even number of photons leads to the decision that $\delta=0$, and of an odd number --- that $\delta\neq0$. Taking into account that the case of $\delta=0$ always gives the ``negative'' result, the corresponding ``false positive'' error probability is equal to zero:
\begin{equation}
  P_{\rm f.p.}  = 0 \,.
\end{equation}
At the same time, the remaining even photon number components in $\ket{\Psi_\delta}$ give the non-vanishing ``false negative'' error probability, equal to
\begin{equation}
  P_{\rm f.n.} = p_{\rm even} = \frac{1+P_\delta}{2} \,.
\end{equation}
The minimum of this probability coincides with the minimum of the parity $P_\delta$. The corresponding value of $\phi$ that provides this minimum approximately equal to \eqref{phi_opt_approx}.

\begin{figure}
	\centering
	\includegraphics[width=0.5\linewidth]{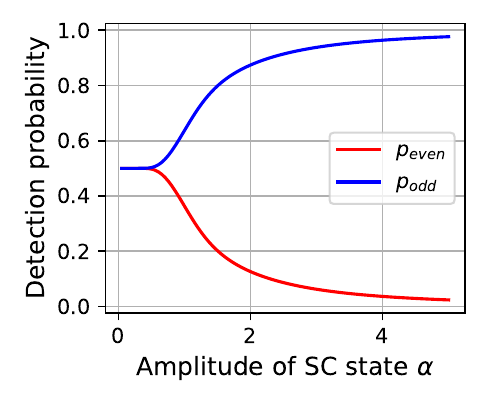}
	\caption{Dependencies of probabilities \(p_{\rm odd}\) and \(p_{\rm even}\) (see Eq.\,\ref{p_even_odd}) on SC state amplitude $\alpha$. Displacement $\delta$ corresponds to the minimum parity of displaced SC state.}
	\label{fig:even-and-odd-probability-ideal-detector}
\end{figure}

In Fig.\,\ref{fig:even-and-odd-probability-ideal-detector}, the probabilities \eqref{p_even_odd} are plotted as functions of \(\alpha\), assuming the optimized values of \(\delta\). It is easy to see for reasonably large values of $\alpha\gtrsim2$, the ``false negative'' probability does not exceed $\sim0.1$.

\subsection{With optical losses}\label{sec:nonideal_SC}

The final state of the optical field  in presence of the optical losses is calculated in Appendix \ref{app:SC_parity}. It is shown there, specifically, that the probability to obtain the even and the odd photon numbers are still defined by Eqs.\,\eqref{p_even_odd}, but with the parity factor equal to
\begin{equation}\label{eq:P_delta_SC_loss}
  P_\delta  = \frac{2e^{-2\delta'^2}}{K}
    (e^{-2\epsilon^2\alpha'^2}\cos4\alpha'\delta' + e^{-2\alpha'^2}) \,,
\end{equation}
where
\begin{subequations}
  \begin{gather}
    \alpha' = \sqrt{\eta}\alpha \,,\quad \delta' = \sqrt{\eta}\delta \,, \\
    \epsilon = \sqrt{\frac{1-\eta}{\eta}} \,.
  \end{gather}
  \label{eq:alpha-beta-prime-epsilon-defenition}
\end{subequations}
In particular, in the absence of the signal, $\delta=0$, the parity factor is equal to
\begin{equation}\label{P_delta_SC_loss_0}
  P_0 = \frac{2}{K}(e^{-2\epsilon^2\alpha'^2} + e^{-2\alpha'^2}) \,,
\end{equation}

\begin{figure}
	\centering
	\includegraphics[width=0.5\linewidth]{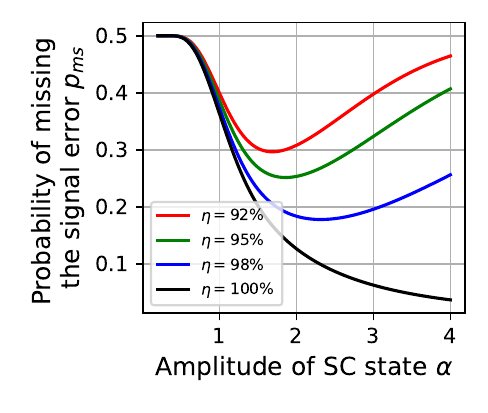}
	\caption{Probability of false detection error for different quantum efficiency of PNR detector \(\eta\) (see Eq.\,\eqref{eq:fd-error-prob})}\label{fig:sc-fd_error_prob}
\end{figure}

It follows from this result, that similarly to the Fock state case, there are several ways how the sensitivity is affected by the optical losses. Firstly, they reduce the effective signal $\delta$ and the amplitude of the SC state by the factor $\sqrt{\eta}$. In case of $1-\eta\ll1$, this problem can be solved by the proportional increase of the pump power or the squeeze factor. Secondly, note that now $P_0<1$, creating the ``false positive'' (false detection) error probability
\begin{equation} \label{eq:fd-error-prob}
  P_{\rm f.p.} = \frac{1-P_0}{2}
  = \frac{1}{K}(1- e^{-2\epsilon\alpha'{}^2})(1 - e^{-2\alpha'{}^2})
\end{equation}
This probability is plotted in Fig.\,\ref{fig:sc-fd_error_prob} for several values of the quantum efficiency $\eta$.

\begin{figure}
	\centering
	\includegraphics[width=0.5\linewidth]{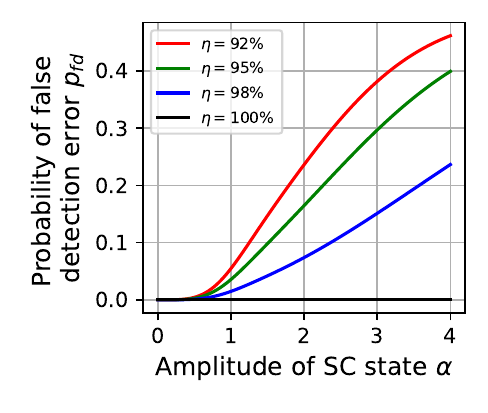}
	\caption{Probability of missing signal error for different quantum efficiency of PNR detector \(\eta\) (see Eq.\,\eqref{eq:sc-ms-error-prob})}\label{fig:sc-ms_error_prob}
\end{figure}

An finally, the losses introduce the additional factor $e^{-2\epsilon^2\alpha'^2}$ in $P_\delta$, thus, as well, increase the ``false negative'' (missing the signal) probability:
\begin{equation}\label{eq:sc-ms-error-prob}
	P_{\rm f.n.} = \frac{1+P_\delta}{2}, \, \delta \neq 0 \,.
\end{equation}
The result is plotted in Fig.\,\ref{fig:sc-ms_error_prob} for several values of the quantum efficiency $\eta$.

	\section{Discussion}\label{sec:conclusion}

The considered scheme of the non-Gaussian interferometer here crucially depends on two additional elements: (i) a deterministic source of light pulses prepared in specific non-Gaussian quantum state, such as Fock or Shr\"odinger cat, and (ii) a highly efficient photon-number-resolving photodetector.

Technologies of preparation of non-Gaussian quantum states have been actively developed in the recent decades, driven by the requirements of quantum information science, see \eg the review article \cite{Lvovsky2020} and the references therein. For example, the fidelity of preparing a single-photon Fock state has reached 76\% \cite{H_Ogawa_PRL_116_23_233602_2016}. The preparation of Schrödinger cat states with \(\alpha^2\sim 3\) was demonstrated with fidelity 77\% \cite{Sychev_NPhot_11_379_2017}. Recently, methods for preparing larger Schrödinger cat states with \(\alpha \gtrsim 4\) and fidelities 95-98\% were proposed \cite{Kuts_PS_97_115002_2022, Podoshvedov_SR_13_3965_2023}.

Concerning the detection, photon-number-resolving detectors with quantum efficiencies up to 98\% have been demonstrated using cryogenic transition-edge sensors \cite{Lita_OE_16_3032_2008, Fukuda_OE_19_870_2011, Gerrits_OE_20_23798_2012, Stasi_PRAppl_19_064041_2023}. This efficiency corresponds to mixing of the non-Gaussian state with a few percents of the vacuum state, promising the sensitivity almost one by order of magnitude better that provided by the coherent quantum state.

Therefore, it is reasonable to assume that the practical implementation of the scheme discussed in this work can be considered feasible.

	\acknowledgments

	This work was supported by the Theoretical Physics and Mathematics Advancement Foundation ``BASIS'' Grant \#23-1-1-39-1.

	The authors would like to thank B.\,Nugmanov for the useful remarks.

\appendix

\section{Derivation of Eq.\,\eqref{Fock_orthog}}\label{app:Fock_orthog}

Note that
\begin{equation}
  \mathcal{D}(\delta) = e^{i\delta\hat{a}^\dagger}e^{i\delta\hat{a}}e^{-\delta^2/2} \,.
\end{equation}
Therefore,
\begin{multline}
	\bra{n}\hat{\mathcal{D}}(\delta)\ket{n}
	=
	\bra{n}e^{i\delta\hat{a}^\dagger}e^{i\delta\hat{a}}\ket{n}e^{-\delta^2/2}
	=
	\bra{n}\sum_{k,l=0}^n\frac{(i\delta\hat{a})^k}{k!}
    \frac{(i\delta\hat{a}^\dag)^l}{l!}\ket{n}e^{-\delta^2/2}
    = \\ =
    \sum_{k=0}^n\frac{(-\delta^2)^k}{(k!)^2}\bra{n}\hat{a}^k\hat{a}^\dag{}^k\ket{n} e^{-\delta^2/2}
    =
 	\sum_{k=0}^n\frac{n!}{(k!)^2(n-k)!}(-\delta^2)^ke^{-\delta^2/2}
 	=
 	\mathcal{L}_n(\delta^2) e^{-\delta^2/2} \,,
\end{multline}
where $\mathcal{L}_n$ is the $n$-th Laguerre polynomial.

\section{Derivation of Eq.\,\eqref{eq:P_delta_SC_loss}}\label{app:SC_parity}

Let $\hat{\mathcal{L}}$ be the unitary operator, describing the optical losses  according to the imaginary beamsplitter model of Ref.\,\cite{Leonhardt_PRA_48_4598_1993}:
\begin{equation}
  \forall\ \alpha: \hat{\mathcal{L}}\ket{\alpha}\ket{0}_B
  = \ket{\sqrt{\eta}\alpha}\ket{\sqrt{1-\eta}\,\alpha}_B \,.
\end{equation}
where the subscript $B$ corresponds to quantum states of the heatbath. In this case, assuming that $\ket{\Psi}_\delta$ is given by Eq.\,\eqref{Psi_delta_SC}, we obtain:
\begin{equation}
  \hat{\mathcal{L}}\ket{\Psi_\delta}\ket{0}_B
  = \frac{1}{\sqrt{K}}\bigl[
        e^{i\alpha\delta}\ket{\alpha'+i\delta'}\ket{\epsilon(\alpha'+i\delta')}_T
        + e^{-i\alpha\delta}\ket{-\alpha'+i\delta'}\ket{\epsilon(-\alpha'+i\delta')}_T
      \bigr] ,
\end{equation}
where the factor $\alpha'$, $\beta'$ and $\epsilon$ are given by Eqs.\,\eqref{eq:alpha-beta-prime-epsilon-defenition}.

Then, trace out the heat bath:
\begin{multline}
  \hat{\rho}_\delta
  = \Tr_T
  	\Big(
        \hat{\mathcal{L}}\ket{\Psi_\delta}\ket{0}_T\,
          {}_T\bra{0}\bra{\Psi_\delta}\mathcal{L}^\dag
      \Big)
  = \frac{1}{K}\Big(
        \ket{\alpha'+i\delta'}\bra{\alpha'+i\delta'}
        + \ket{-\alpha'+i\delta'}\bra{-\alpha'+i\delta'} \\
        + e^{-2\epsilon^2\alpha'^2}[
              e^{2i\alpha'\delta'}\ket{\alpha'+i\delta'}\bra{-\alpha'+i\delta'}
              + e^{-2i\alpha'\delta'}\ket{-\alpha'+i\delta'}\bra{\alpha'+i\delta'}
            ]
      \Big) .
\end{multline}
The corresponding probability distribution for $n$ is equal to
\begin{equation}\label{P_n_loss}
  p_n = \bra{n}\hat{\rho}_\delta\ket{n}
  = \frac{2e^{-\alpha'^2-\delta'^2}}{Kn!}\Bigl(
        (\alpha'^2+\delta'^2)^n
        + e^{-2\epsilon^2\alpha'^2}
            \Re\bigl\{e^{2i\alpha'\delta'}[-(\alpha'+i\delta')^2]^n\bigr\}
      \Bigr) .
\end{equation}

Now, using the summation rules
\begin{equation}\label{sums}
  \cosh x = \sum_{n=0}^\infty\frac{x^{2n}}{(2n)!} \,, \quad
  \sinh x = \sum_{n=0}^\infty\frac{x^{2n+1}}{(2n+)!}
\end{equation}
we can calculate the probabilities for the even and odd photon numbers:
\begin{equation}
  p_{\rm even} = \sum_{n=0}^\infty p_{2n} \,, \quad
  p_{\rm odd} = \sum_{n=0}^\infty p_{2n+1} \,.
\end{equation}
It is easy to see that they are again defined by Eqs.\,\eqref{p_even_odd}, but with the parity factor equal to \eqref{eq:P_delta_SC_loss}.


%

\end{document}